\documentclass[twocolumn,english,superscriptaddress]{revtex4-2}

\usepackage{mathpazo}
\usepackage[T1]{fontenc}
\usepackage[latin9]{inputenc}
\usepackage{color}
\usepackage{babel}
\usepackage{amsmath}
\usepackage{amssymb}
\usepackage{graphicx}
\usepackage[unicode=true,
 bookmarks=false,
 breaklinks=true,pdfborder={0 0 1},backref=false,colorlinks=true,pdfpagemode=FullScreen]
 {hyperref}
\hypersetup{
 pdfborderstyle=,citecolor=blue,filecolor=black,linkcolor=red,urlcolor=blue}

\begin{document}

\title{Spontaneous symmetry breaking induced by nonlinear interaction in
a coupler supported by fractional diffraction}

\author{Mateus C. P. dos Santos}
\email{mateuscalixtopereira@gmail.com}
\address{Instituto de Ciências Tecnológicas e Exatas, Universidade Federal do Triângulo Mineiro, 38066-200, Uberaba, Brazil}
\address{Federal Institute of Maranhão, PPGCTM, 65030-005, São Luís, Maranhão,
Brazil}

\author{Wesley B. Cardoso}
\email{wesleybcardoso@ufg.br}
\address{Instituto de Física, Universidade Federal de Goiás, 74.690-970, Goiânia,
Goiás, Brazil}

\begin{abstract}
In this paper we introduce a one-dimensional model of coupled fractional
nonlinear Schrödinger equations with a double-well potential applied
to one component. This study examines ground state (GS) solitons,
observing spontaneous symmetry breaking (SSB) in both the actuated
field and the partner component due to linear coupling. Numerical
simulations reveal symmetric and asymmetric profiles arising from
a slightly asymmetric initial condition. Asymmetry is influenced by
nonlinearities, potential depth, and coupling strength, with self-focusing
systems favoring greater asymmetry. Fractional diffraction affects
the amplitude and localization of symmetric profiles and the stability
of asymmetric ones. We identify critical Lévy index values for
generating coupled GS solitons. Stability analysis of unstable, centrally
asymmetric GS solitons demonstrates oscillatory dynamics, providing
new insights into SSB in fractional systems and half-trapped solitons. 
\end{abstract}
\keywords{Spontaneous symmetry breaking; Lévy index; Nonlinear fractional
Schrödinger equations; Half-trapped systems.}
\maketitle

\section{Introduction \label{Intro}}

Solitons, which are localized wave packets, retain their shape and
stability during propagation as a result of a precise equilibrium
between dispersion and nonlinearity. This phenomenon occurs in a wide
range of physical systems, including optical fibers, plasmas, and
Bose-Einstein condensates (BECs), where nonlinear interactions effectively
counteract the dispersive forces. The formation and stability of solitons
have been extensively explored in nonlinear systems, providing a rich
framework for understanding wave dynamics under nontrivial conditions
\citep{Agrawal_13}. Due to their ability to preserve their structure
and interact with other solitons without losing coherence, solitons
have become a central topic in the study of complex wave phenomena
\citep{Kivshar_03}. Moreover, the analysis of solitons is essential
for advancing the understanding of more intricate nonlinear behaviors,
especially when additional influences, such as fractional diffraction,
are introduced.

Fractional diffraction, which extends classical diffraction by introducing
fractional-order derivatives in the governing equations, brings about
novel dynamics that influence soliton behavior. The application of
the fractional Laplacian operator induces long-range interactions
and anomalous diffusion, significantly complicating the solitons'
stability, localization, and interaction properties \citep{Longhi_OL15,Malomed_P21}.
These nonlocal effects can reshape soliton propagation, often creating
conditions that promote the onset of more intricate phenomena, such
as spontaneous symmetry breaking (SSB). The study of fractional diffraction
and its impact on soliton dynamics has gained considerable attention,
providing deeper insights into wave propagation in complex and nonlocal
media \citep{Yao_PR18,Dong_OE18,Zhu_OE20,Li_OL21,Che_PLA21,Wu_CSF22,Bo_ND23,Zhong_CP23,Wang_PLA23,Chen_PD24,Wang_PD24,Yu_PD24,Bai_PS24,Zhai_AOS24}.

Indeed, SSB plays a crucial role in physics, describing how a system
initially in a symmetric configuration evolves into an asymmetric
state due to internal dynamics or external perturbations \citep{Kevrekidis_08}.
In the context of coupled nonlinear systems, SSB can give rise to
asymmetric soliton states, even when the governing equations retain
symmetry. The interplay between SSB and fractional diffraction is
especially compelling, as the nonlocal effects induced by fractional
diffraction can shift the power balance between coupled components,
leading to symmetry disruption and the formation of localized structures
\citep{Zeng_PD23,He_CSF24}.

Another interesting effect is the half-trapped solitons, which emerge
in systems governed by coupled nonlinear Schr{ö}dinger equations
(NLSE), where a confining potential is applied to only one of the
fields, while the second (partner) field remains unconfined \citep{Hacker_S21}.
In such configurations, localization occurs in the unconfined field
solely due to the coupling between the two fields, as the second field
would otherwise remain delocalized in the absence of this interaction.
The coupling induces a mutual localization effect, allowing the partner
field to inherit the confinement properties of the first, resulting
in a spatially localized soliton state that would not be achievable
without the coupled dynamics. In this sense, Ref. \citep{dosSANTOS_PRE21}
has presented the study on Anderson localization induced by interaction
in linearly coupled binary BECs. Also, the phenomenon of SSB induced
by interaction in linearly coupled binary BECs has been previously
reported in Ref. \citep{Santos_ND23}. The objective of the present
work is to investigate how linear coupling induces spontaneous symmetry
breaking (SSB) in a system characterized by fractional diffraction.
Specifically, this study aims to explore the interplay between linear
coupling and fractional diffraction effects, examining how these interactions
influence the emergence and characteristics of SSB in such a nonlinear
system.

The structure of the paper is as follows: The subsequent section introduces
the theoretical model. Section \ref{Sec3} is dedicated to the numerical
simulations and the analysis of their results. Finally, Section \ref{Sec4}
presents our conclusions.

\section{Theoretical model \label{Sec2}}

We begin the analysis by considering the propagation of light in a
system composed of coupled planar waveguides, described by the rescaled
model of the coupled nonlinear fractional Schr{ö}dinger equation
(f-NLSE) \citep{Zeng_CHAOS20,Li_CHAOS22,Ur_OQE22,Ahmad_RP23} 
\begin{align}
i\dfrac{\partial}{\partial z}\phi_{1} & =\frac{1}{2}\left[-\dfrac{\partial^{2}}{\partial x^{2}}\right]^{\alpha/2}\phi_{1}+\left(g_{1}|\phi_{1}|^{2}+g_{12}|\phi_{2}|^{2}\right)\phi_{1}\nonumber \\
 & \quad+U(x)\phi_{1},\nonumber \\
i\dfrac{\partial}{\partial z}\phi_{2} & =\frac{1}{2}\left[-\dfrac{\partial^{2}}{\partial x^{2}}\right]^{\alpha/2}\phi_{2}+\left(g_{2}|\phi_{2}|^{2}+g_{12}|\phi_{1}|^{2}\right)\phi_{2},\label{EQ1}
\end{align}
where $z$ represents the normalized propagation distance, $x$ denotes
the transverse coordinate, and $\alpha$ is the L{é}vy index (LI).
The functions $\phi_{1,2}(x,z)$ describe the field amplitudes corresponding
to components (waveguides) 1 and 2, respectively. The parameters $g_{1,2}$
represent the Kerr self-interaction coefficients, while $g_{12}$
denotes the nonlinear interaction coefficient between the components.
The sign of the nonlinearity parameters, $g_{1,2}<0$ or $g_{1,2}>0$,
characterizes the system as exhibiting either self-focusing or self-defocusing
nonlinearity, respectively. Model (\ref{EQ1}) describes a conservative
optical system. In this context, the dynamically invariant quantity
of total power (or norm) is expressed as 
\begin{equation}
P\equiv P_{1}+P_{2}=\int_{-\infty}^{+\infty}\left(|\phi_{1}|^{2}+|\phi_{2}|^{2}\right)dx,\label{NORM-1}
\end{equation}
where $P_{1}$ and $P_{2}$ are the individual powers of each component.

The LI $\alpha$, which ranges from $0<\alpha\leq2$, is featured
in the fractional diffraction operator $(-\partial{^{2}}/\partial x{^{2}})^{\alpha/2}$,
representing the kinetic energy term in the Schr{ö}dinger equation.
When $\alpha=2$, this operator reduces to the standard 1D Laplacian,
resulting in the coupled NLSE with quadratic diffraction in Eq. (\ref{EQ1}).
By using the direct and inverse Fourier transform, the fractional
Laplacian in Eq. (\ref{EQ1}) can be rewritten in terms of the integral
operator \citep{Laskin_PRE00,Laskin_PLA00} 
\begin{align}
\left(-\dfrac{\partial^{2}}{\partial x^{2}}\right)^{\alpha/2}\phi(x,z) & =\frac{1}{2\pi}\iint dkdx'|k|^{\alpha}\nonumber \\
\times & \exp\left[ik(x-x')\right]\phi(x',z).\label{lag}
\end{align}
In the self-focusing f-NLSE, which can be derived from the equation
of motion for the component $\phi_{2}$ with $g_{2}<0$ and $g_{1,2}=0$
(see Eq. (\ref{EQ1})), collapse occurs when $\alpha\leq1$, inhibiting
the formation of stable localized solutions \citep{malomed_rev_P21}.

In this work we consider an external potential acting only on component
1, which has the function of spatially confining the respective optical
field. Systems with two components, where only one of them is trapped,
have been previously studied in linearly coupled NLSE. In these cases,
the so-called half-trapped systems exhibit ground state and dipole
modes under harmonic oscillator-type confinement \citep{HACKER_SYMMETRY21},
and Anderson localization when a quasi-periodic optical lattice is
considered \citep{dosSANTOS_PRE21}. In this context, to induce symmetry
breaking, we employ a double-well potential, defined by

\begin{equation}
U(x)=-U_{0}\left[\cosh(x+x_{0})^{-2}+\cosh(x-x_{0})^{-2}\right],\label{Pot}
\end{equation}
where $U_{0}$ denotes the potential depth, and $x_{0}$ specifies
the position of the local minimum for each of the wells. For simplicity,
in our simulations we will assume $x_{0}=3$. Similar potentials have
been utilized to generate asymmetric states in the study of ultracold
gases, where a nonpolynomial Schr{ö}dinger equation (NPSE), derived
from a variational approach, is employed to model the longitudinal
dynamics of a BEC confined in the transverse direction \citep{Miranda_PLA22}.
Also, in Ref. \citep{Santos_ND23}, the induction of SSB in a half-trapped
system described by a linearly coupled NLSE was achieved through a
double-well potential (similar to Eq. (\ref{Pot})) in framework of
binary BECs. Unlike previous works, this study proposes investigating
the induction of SSB in fractional half-trapped optical systems with
nonlinear interaction, which is generally present in high-intensity
optical field configurations. 
\begin{figure}[tb]
\centering \includegraphics[width=1\columnwidth]{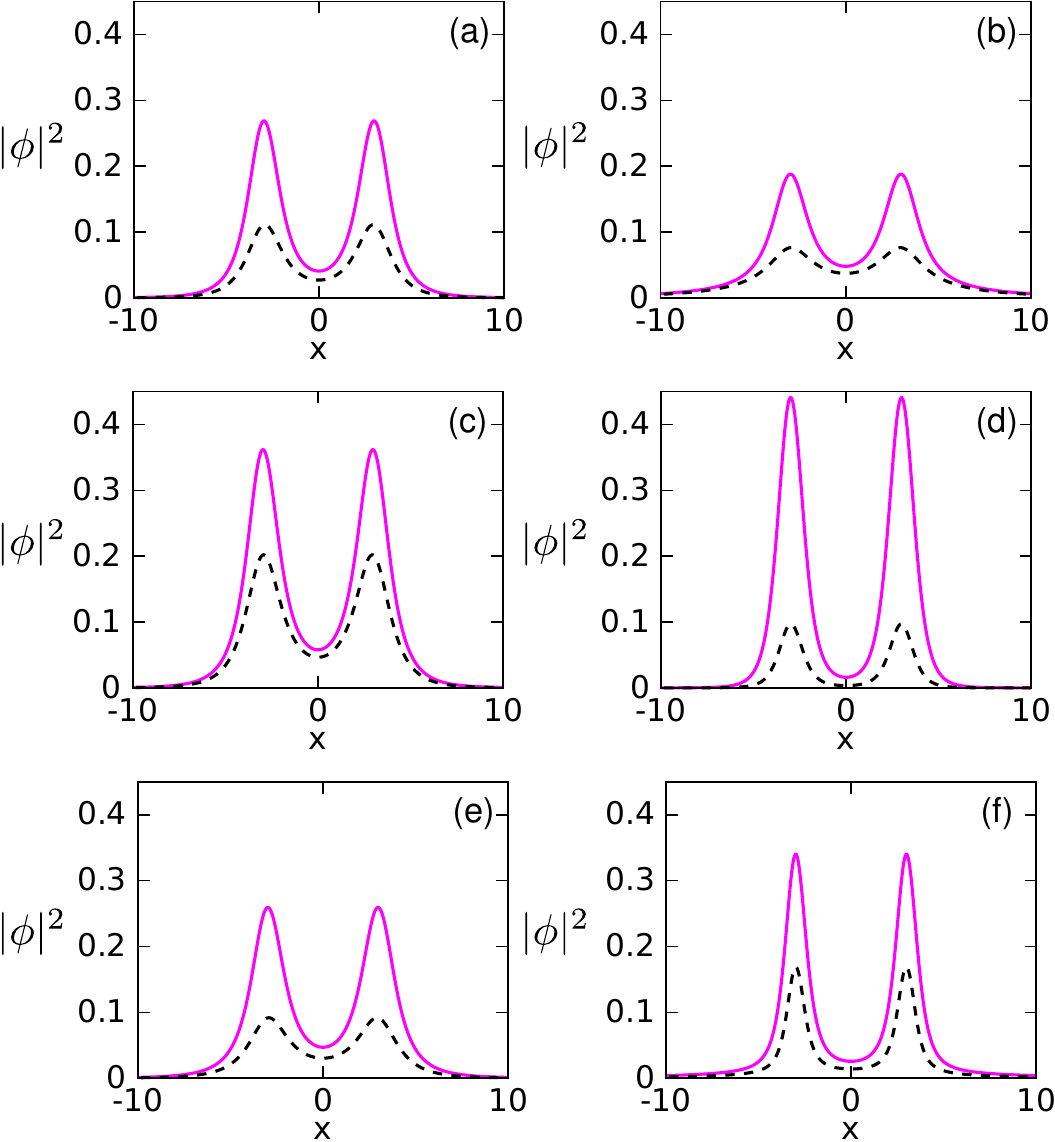}

\caption{Symmetric profiles of the coupled states. The components $|\phi_{1}(x)|^{2}$
and $|\phi_{2}(x)|^{2}$ which were obtained numerically by Eq. (\ref{EQ1})
are arranged in magenta solid lines and black dotted lines. The parameters
used in (a) are: $P=2$, $U_{0}=1$, $g_{1}=g_{2}=3$, $g_{12}=-3$
and $\alpha=1.5$. The other panels are obtained with the same parameters
as in (a), but for (b) $g_{1}=g_{2}=5$, (c) $P=3$, (d) $U_{0}=2$,
(e) $g_{12}=-2.5$, and (f) $\alpha=0.5$.}

\label{F1} 
\end{figure}

\begin{figure*}[t]
\centering \includegraphics[width=0.8\textwidth]{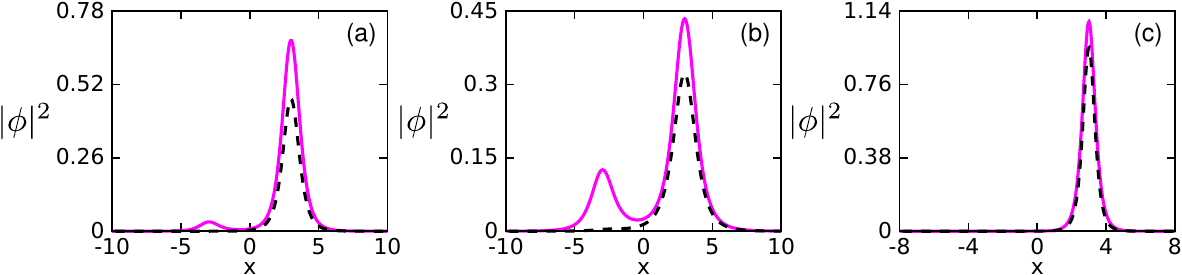} \caption{Asymmetric profiles of the coupled states $|\phi_{1}(x)|^{2}$ and
$|\phi_{2}(x)|^{2}$ versus $x$, obtained in the same configurations
as in Fig. \ref{F1}(a), but for $g_{1}=g_{2}=2$ (a), $g_{2}=2$
(b), and $g_{12}=-5$ (c).}
\label{F2} 
\end{figure*}

\begin{figure*}[t]
\centering \includegraphics[width=0.9\textwidth]{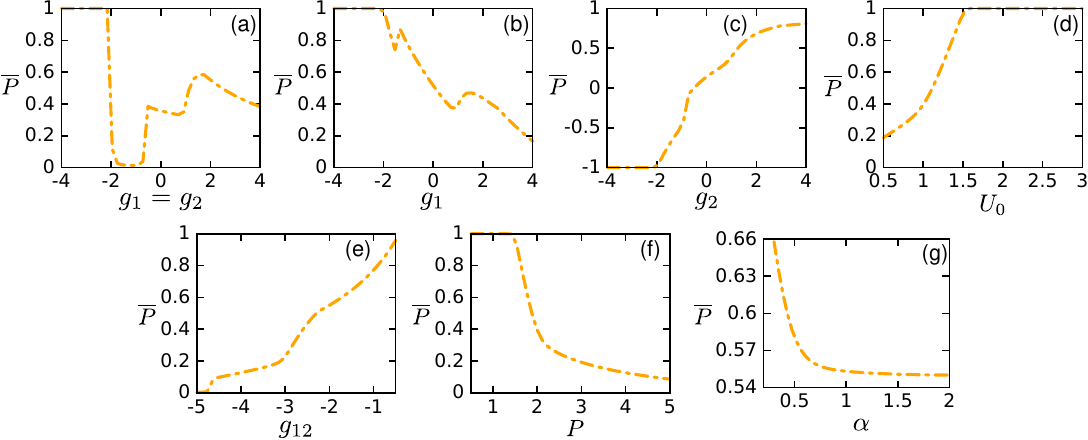} \caption{Comparison between the individual powers $P_{1}$ and $P_{2}$, considering
the relative power $\overline{P}=\left(P_{1}-P_{2}\right)/P$ versus
the parameters (a) $g_{1}=g_{2}$, (b) $g_{1}$, (c) $g_{2}$, (d)
$U_{0}$, (e) $g_{12}$, (f) $P$ and (g) $\alpha$. The parameters
used here are: (a) $P=2$, $U_{0}=1$, $\alpha=1.5$ and $g_{12}=-2$;
(b) the same as in (a) but for $g_{2}=1$; (c) the same as in (a)
but for $g_{1}=1$; (d) $P=2$, $g_{1}=g_{2}=2$, $\alpha=1.5$, and
$g_{12}=-2$; (e) the same as in (d) but for $U_{0}=1$; (f) the same
as in (d) but for $U_{0}=1$; and (g) the same as in (d).}
\label{F3} 
\end{figure*}

\begin{figure*}[t]
\centering \includegraphics[width=0.8\textwidth]{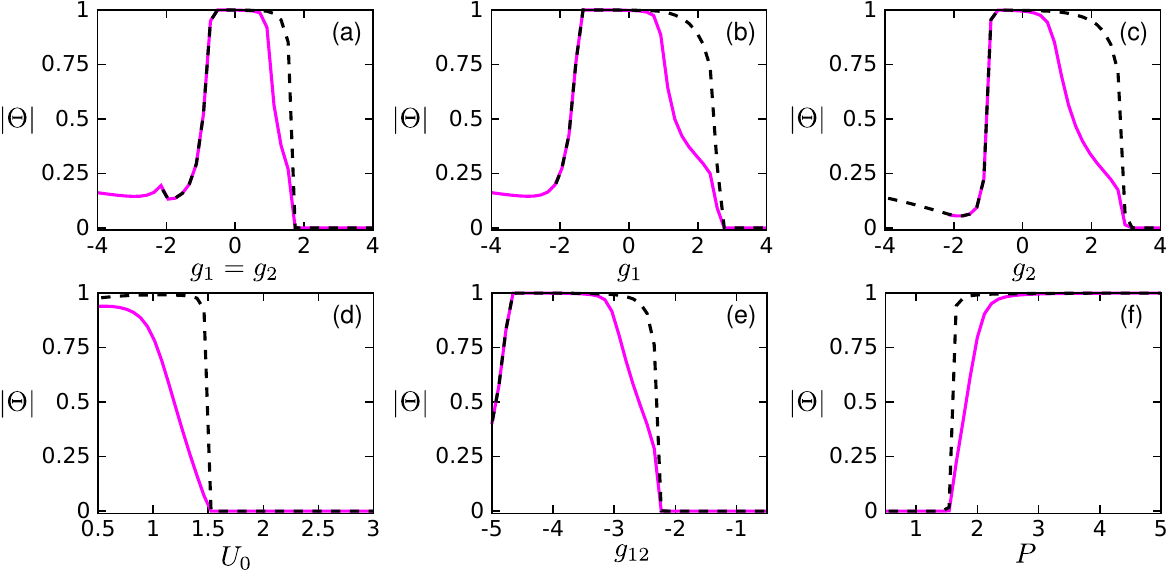} \caption{Absolute values of the asymmetry rate $|\Theta|$ versus (a) $g_{1}=g_{2}$,
(b) $g_{1}$, (c) $g_{2}$, (d) $U_{0}$, (e) $g_{12}$, and (f) $P$,
considering the same settings as the respective panels in Fig. \ref{F3}.
The results of the components $\phi_{1}$ and $\phi_{2}$, corresponding
to $|\Theta_{1}|$ and $|\Theta_{2}|$ are shown in solid magenta
lines and dotted black lines, respectively.}
\label{F4} 
\end{figure*}

\begin{figure}[tb]
\centering \includegraphics[width=1\columnwidth]{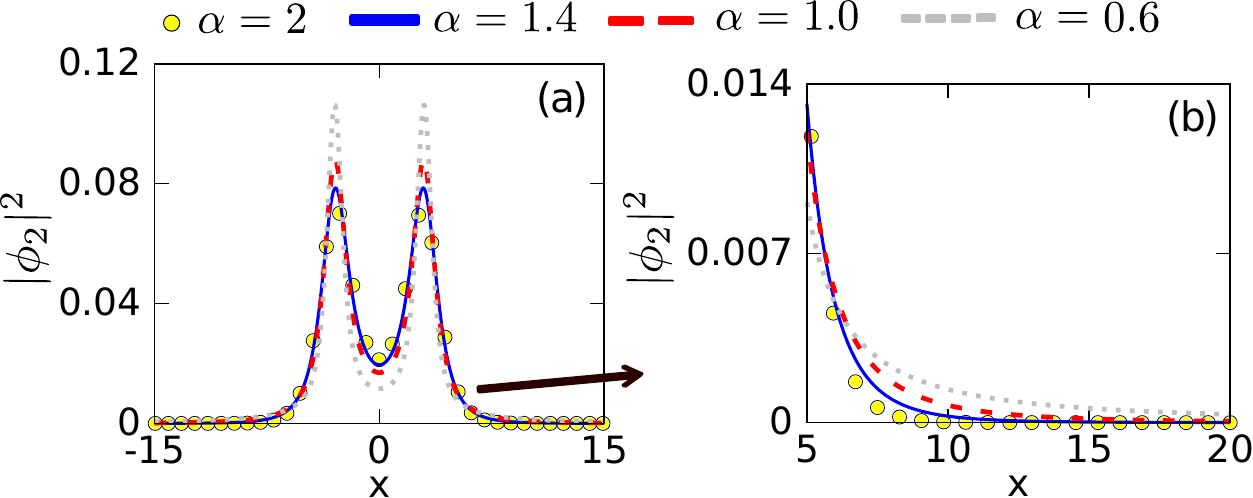}

\caption{Effect of fractional diffraction ($\alpha$) on the symmetric coupled
state of component 2 ($\phi_{2}$). The profiles were obtained numerically
through Eq. (\ref{EQ1}) with $g_{1}=g_{2}=2$, $g_{12}=-2$, $U_{0}=1$
and $P=2$.}

\label{F5} 
\end{figure}

\begin{figure}[tb]
\centering \includegraphics[width=1\columnwidth]{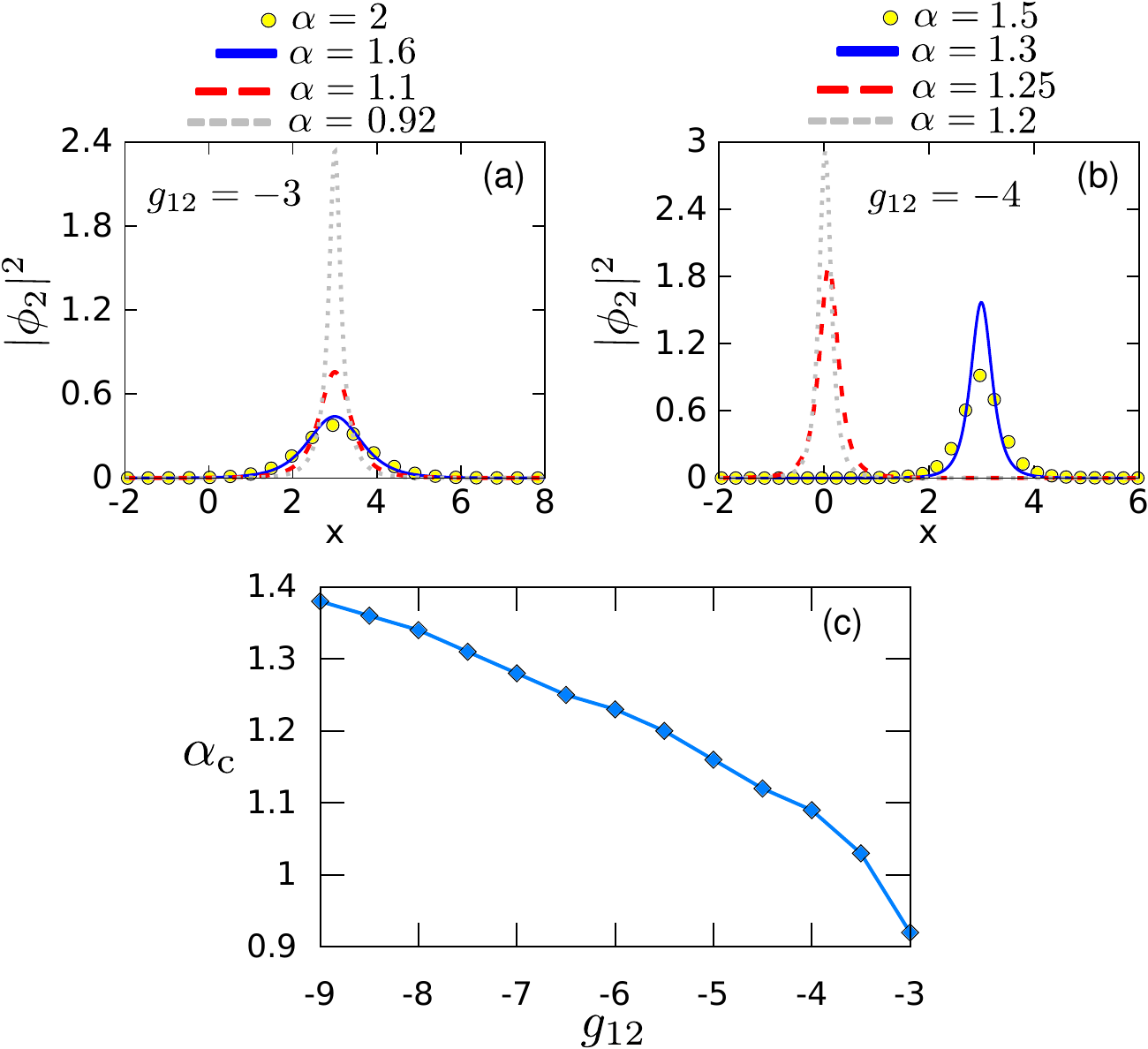}

\caption{The same as Fig. \ref{F5}, but for asymmetric coupled state of component
2 ($\phi_{2}$) with (a) $g_{12}=-3$ and (b) $g_{12}=-4$. The asymmetric
profiles have $\Theta_{2}(\alpha=1.5)=1$, $\Theta_{2}(\alpha=1.3)=1$,
$\Theta_{2}(\alpha=1.25)=0.33$ and $\Theta_{2}(\alpha=1.2)=0.13$.
(c) Critical values of LI ($\alpha_{c}$) versus $g_{12}$, considering
$g_{1}=g_{2}=2$, $U_{0}=1$ and $P=2$. Only singular states are
obtained with $\alpha<\alpha_{c}$.}

\label{F6} 
\end{figure}

\begin{figure*}[t]
\centering \includegraphics[width=0.8\textwidth]{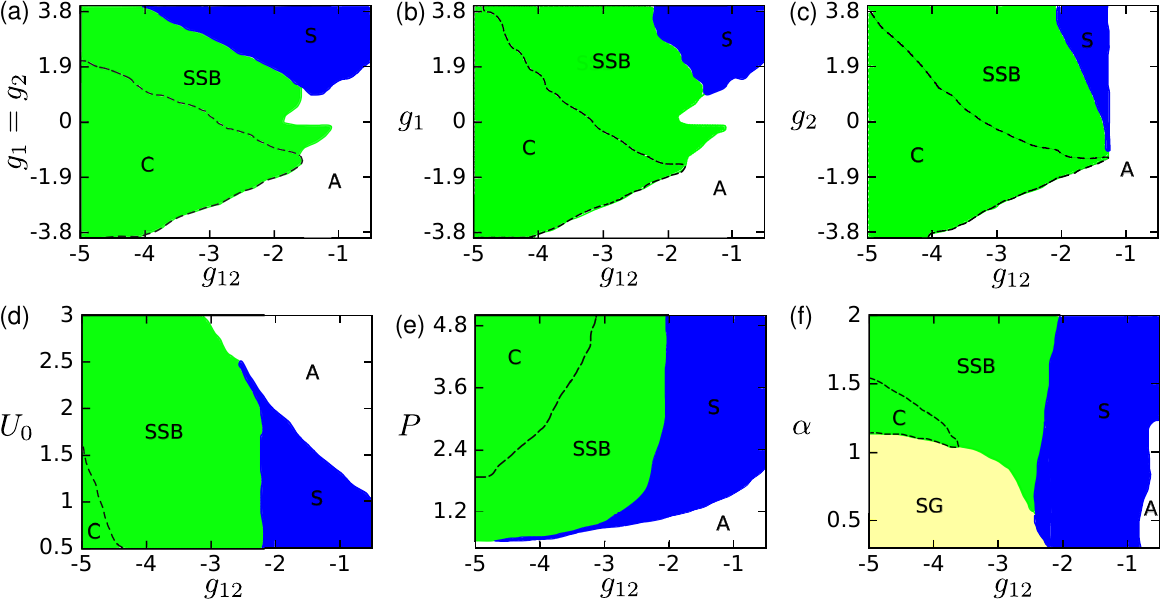} \caption{Phase diagram of coupled states. The symmetric phase (S) is represented
by the blue filled area while spontaneous symmetry breaking (SSB)
is represented by the green filled area. The empty areas (A) represent
the regions where coupled states are not allowed ($P_{1}=0$ or $P_{2}=0$).
Unstable delta-type configurations (SG) are represented by the yellow
filled area, while the region enclosed by the black dashed line indicates
the central asymmetric configurations (C). The other parameters used
here are: (a) $P=2$, $U_{0}=1$ and $\alpha=1.5$; (b) $P=2$, $U_{0}=1$,
$g_{2}=1$ and $\alpha=1.5$; (c) $P=2$, $U_{0}=1$, $g_{1}=1$ and
$\alpha=1.5$; (d) $P=2$, $g_{1}=g_{2}=2$ and $\alpha=1.5$; (e)
$U_{0}=1$, $g_{1}=g_{2}=2$ and $\alpha=1.5$; and (f) $P=2$, $U_{0}=1$
and $g_{1}=g_{2}=2$.}
\label{F7} 
\end{figure*}

\begin{figure*}[t]
\centering \includegraphics[width=0.8\textwidth]{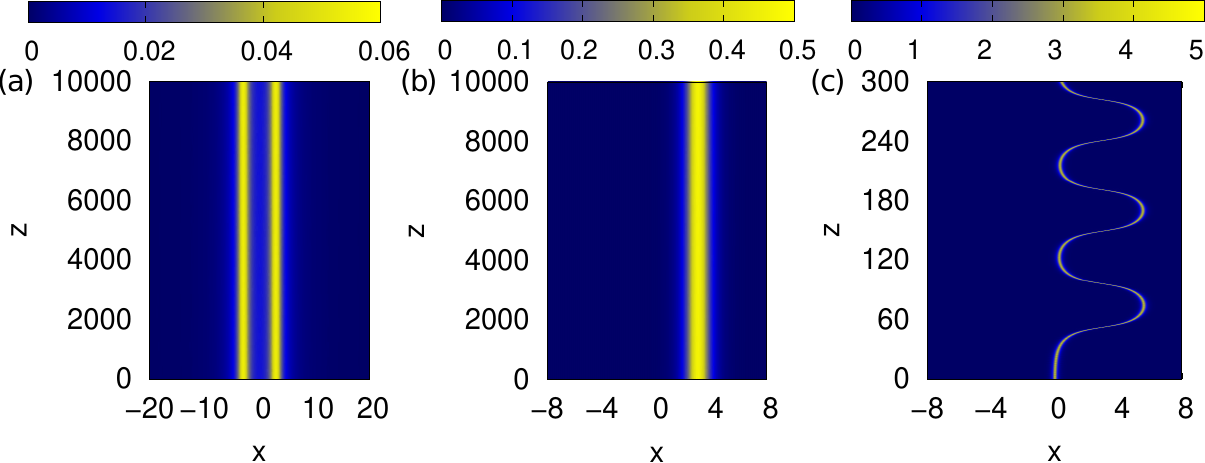} \caption{Evolution of stable GS solitons (a) symmetric and (b) asymmetric.
(c) Typical unstable evolution of a central asymmetric profile. In
all panels, only the results for the field density $|\phi_{2}|^{2}$
are presented. All results presented refer to component 2. The other
parameters used here are: $U_{0}=1$, $P=2$, $g_{1}=g_{2}=1$ (a)
$\alpha=0.9$ and $g_{12}=-1.5$; (b) $\alpha=1.5$ and $g_{12}=-3$;
and (c) $\alpha=1.3$ and $g_{12}=-4.5$.}
\label{F8} 
\end{figure*}

\begin{figure*}[t]
\centering \includegraphics[width=0.8\textwidth]{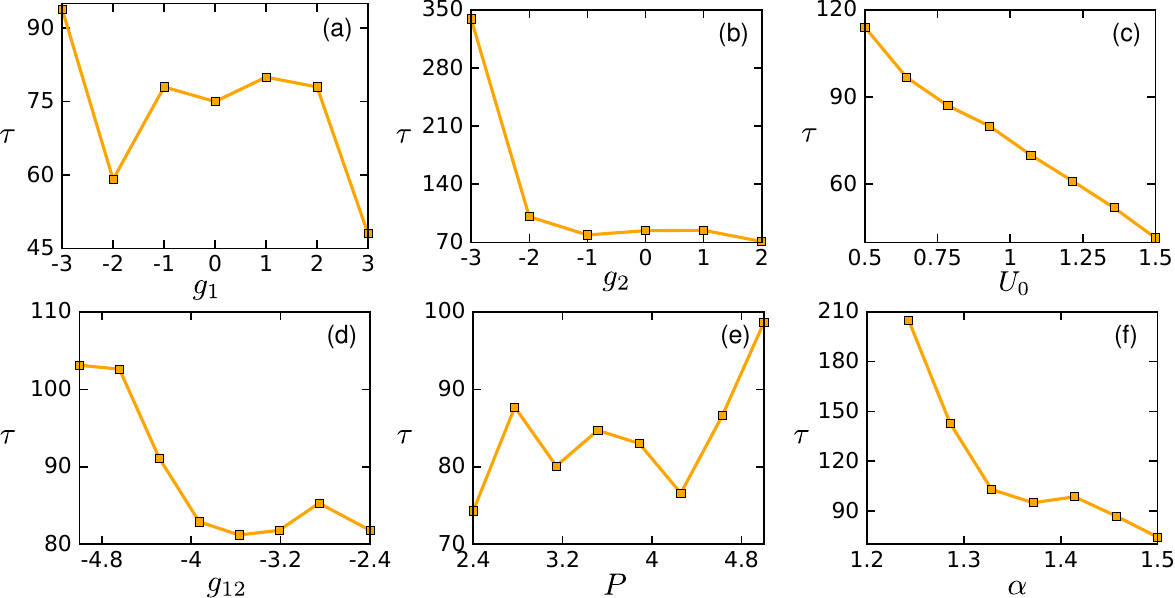} \caption{Oscillation periods of the unstable central asymmetric solitons corresponding
to component 2. The parameters used are: (a) $P=2$, $U_{0}=1$, $\alpha=1.5$,
$g_{2}=1$ and $g_{12}=-4.5$; (b) the same as in (a) but for $g_{2}=1$;
(c) $P=2$, $U_{0}=1$, $\alpha=1.5$, $g_{1}=g_{2}=2$ and $g_{12}=-5$;
(d) $P=2$, $U_{0}=1$, $\alpha=1.5$, and $g_{1}=g_{2}=-1.5$; (e)
$U_{0}=1$, $\alpha=1.5$, $g_{1}=g_{2}=2$ and $g_{12}=-4.5$; and
(f) $P=2$, $U_{0}=1$, $g_{12}=-5$, and $g_{1}=g_{2}=-1.5$.}
\label{F9} 
\end{figure*}

\section{Numerical simulations \label{Sec3}}

Deriving analytical solutions for coupled f-NLSEs is particularly
challenging due to the interplay of nonlinearity, coupling, and fractional
diffraction. Recent studies indicate that the inclusion of the fractional
Laplacian operator $(-\partial{^{2}}/\partial x{^{2}})^{\alpha/2}$
significantly complicates the process of obtaining general analytical
solutions \citep{Zhang_PRL15,Zhang_LPR16,Zhong_PRE16,Chen_PRE18,Chen_CNSNS19,Zeng_CHAOS20,Zeng_L_Dimutri_Chaos21,DOSSANTOS2024CSF},
making numerical methods the preferred approach in such scenarios.
Consequently, in this work, solutions are determined numerically by
integrating the equations of motion (\ref{EQ1}) using both imaginary-time
and real-time propagation methods.

The imaginary-time propagation method promotes the ground states (GS)
by replacing the coordinate $z\rightarrow-iz$ in Eqs. (\ref{EQ1}).
In this sense, the propagation in imaginary ``time'' (respective
of the propagation coordinate $z$) acts as a relaxation process over
an input profile $\phi_{1,2}^{\textrm{In}}(x,z=0)$, whose result
converges to the lowest energy state related to the configuration
of Eq. (\ref{EQ1}). To find the GS solitons, we start the simulations
with the slightly asymmetric Gaussian input profile

\begin{equation}
\phi_{1,2}^{\textrm{In}}(x,z=0)=\mathcal{P}\exp\left[-\frac{(x-\sigma)^{2}}{2}\right],\label{assim}
\end{equation}
where $\mathcal{P}$ is the normalization constant defined according
to the total power (\ref{NORM-1}) and $\sigma=0.01$ is the asymmetry
parameter. During the relaxation process, both individual powers are
free, obeying only the normalization condition (\ref{NORM-1}). Therefore,
the GS solitons exhibit $P_{1,2}=[0,\,P]$, with $P_{1}+P_{2}=P$.
The real-time propagation method is employed to investigate the dynamics
of the GS solitons, which are used as input profiles, from which their
behavior along $z$ is studied. Both numerical methods are performed
by a second-order split-step algorithm used to separately integrate
the linear and nonlinear terms of the system of Eq. (\ref{EQ1}).
The diffractive part $\sim(-\partial{^{2}}/\partial x{^{2}})^{\alpha/2}$
is integrated using Eq. (\ref{lag}), while the nonlinear part presents
an analytical solution. The integration of both terms is performed
simultaneously in each iteration step \citep{Yang_10}.

For self-focusing systems ($g_{1,2}<0$), even in the absence of coupling
($g_{12}=0$), model (\ref{EQ1}) presents GS solitons for both components.
However, when self-defocusing systems are considered ($g_{1,2}>0$),
only component 1 presents GS solutions in the absence of coupling.
Therefore, in this configuration, the localization of component 2
occurs only when it is induced by the partner field coupling. Additionally,
the simulations show that the coupling determines the appearance of
localized solutions. For example, considering $g_{12}<0$, the coupling
acts as a confining mechanism, similar to a spatially inhomogeneous
nonlinearity \citep{Zeng_L_Dimutri_Chaos21,Li_CHAOS22,dosSantos_CJP24}.
Conversely, when $g_{12}>0$, the coupling operates as a delocalizing
factor, inhibiting the formation of localized solutions such as GS
solitons.

In the numerical simulations, we found coupled GS solitons that present
spatial symmetry with respect to the $x$-axis. These symmetric GS
solitons present two identical peaks centered at $\pm x_{0}$. In
Fig. \ref{F1}, we studied the influence of nonlinearities ($g_{1},g_{2}$),
total power ($P$), potential depth ($U_{0}$), coupling intensity
($g_{12}$) and LI ($\alpha$) on the coupled symmetrical profiles
obtained in the self-defocusing regime. Specifically, the Fig. \ref{F1}(a)
shows a typical example of symmetrical GS solitons. We observe that
increasing the nonlinearities ($g_{1}=g_{2}$) leads to a broadening
of the profiles and a reduction in amplitude, as shown in Fig. \ref{F1}(b).
The increase in total power also promotes a rise in the amplitude
of both components (see Fig. \ref{F1}(c)). In Fig. \ref{F1}(d),
we present the profiles resulting from the increase in the depth of
the double-well potential. Note that component 1 has a much higher
amplitude than component 2, with peaks that are almost completely
separated from each other. A slight reduction in the coupling intensity
($|g_{12}|$) also promotes the amplitude difference between the components,
as shown in Fig. \ref{F1}(e). Fig. \ref{F1}(f) was produced with
$\alpha=0.5$, which, compared to Fig. \ref{F1}(a), shows that the
reduction in symmetric fractional diffraction causes a narrowing of
the peaks in the GS solitons.

Our numerical analysis reveals the presence of spatially asymmetric
GS solutions, characterized either by peaks of differing amplitudes
or by a single peak displaced from the center of the double-well potential.
Therefore, for certain parameters (detailed below), SSB occurs, resulting
in GS solitons characterized by $|\phi_{1,2}(x,z)|^{2}\neq|\phi_{1,2}(-x,z)|^{2}$.
In Fig. \ref{F2}, we present three examples of asymmetric GS solitons
obtained from the variation of $g_{1}=g_{2}$, $g_{2}$, and $g_{12}$,
considering the configuration shown in Fig. \ref{F1}. It is observed
that reducing the nonlinearities of both components in self-defocusing
systems can induce SSB. In particular, reducing the nonlinearity of
component 2 is sufficient by itself to promote SSB. Furthermore, the
coupling also influences the symmetry type of the coupled GS solitons,
where an increase in $|g_{12}|$ promotes asymmetry. It is important
to highlight that both profiles were obtained in self-defocusing systems,
where the absence of coupling renders the formation of coupled GS
solitons impossible. As a result, the SSB (and localization) observed
in component 2 are induced by component 1.

The asymmetric profiles in Fig. \ref{F2} are generated with $\sigma=0.01$,
and therefore, their highest peaks (or single peak) are shifted in
the direction of $x>0$. Numerical simulations show that the same
asymmetric GS solitons are obtained using $\tilde{\sigma}=-\sigma$,
but with the inversion of the direction of the highest peaks, i.e.,
$|\phi_{1,2}^{\sigma}(x,z)|^{2}=|\phi_{1,2}^{\tilde{\sigma}}(-x,z)|^{2}$.
Therefore, the set of these asymmetric GS solitons obtained with different
symmetry parameters ($\sigma$, $\tilde{\sigma}$) forms degenerate
states of Eq. (\ref{EQ1}).

Based on Figs. \ref{F1} and \ref{F2}, it is evident that the individual
powers of components 1 and 2 can vary depending on the selected parameters.
This behavior is effectively captured by the relative power $\overline{P}=\left(P_{1}-P_{2}\right)/P$,
where $\overline{P}>0$ ($\overline{P}<0$) indicates that $P_{1}>P_{2}$
($P_{2}>P_{1}$), while $\overline{P}=0$ represents equality between
the individual powers. In Fig. \ref{F3}(a), we present the influence
of both nonlinearities ($g_{1}=g_{2}$) on $\overline{P}$. We observe
that in the strong self-focusing regime, $g_{1}=g_{2}<-2$, only component
1 presents a localized solution. This behavior is abruptly modified
for $g_{1}=g_{2}>-2$, where both components present similar values.
In the entire analyzed region, we find $P_{1}\geq P_{2}$. To analyze
the effects of $g_{1}$ on the individual powers of the GS solitons,
we fix the nonlinearity of component 2 ($g_{2}=1$). Similar to the
previous case (see Fig. \ref{F3}(a)), it is observed that in the
strong self-focusing regime the power of component 1 is favored. However,
increasing $g_{1}$ leads $P_{1}\rightarrow P_{2}$, as shown in Fig.
\ref{F3}(b). The same analysis performed for $g_{2}$ produces very
different results. In the configuration where the nonlinearity of
component 1 is fixed at $g_{1}=1$, shown in Fig. \ref{F3}(c), the
decrease of $g_{2}$ favors $P_{2}$. For example, considering $g_{2}<-2$,
only component 2 presents a localized solution.

Fig. \ref{F3}(d) shows the effects of the double-well potential depth
on the individual powers. We observe that increasing $U_{0}$ causes
an imbalance between the individual powers. In this configuration,
all GS solitons obtained with $U_{0}>1.6$ exhibit only component
1 ($P_{1}=P$, $P_{2}=0$). Similarly, reducing the coupling strength
between the components leads to the vanishing of component 2, as observed
in Fig. \ref{F3}(e). From Fig. \ref{F3}(f), we observe that increasing
the total power $P$ results in GS solitons with similar individual
powers. Conversely, under the same conditions, only component 1 is
obtained when $P<1.4$. Unlike the previous analyses, the influence
of fractional diffraction on the individual powers is relatively small.
In general, increasing $\alpha$ induces slight changes in the individual
powers, reducing the imbalance between the components.

Fig. \ref{F3} shows how the individual norms are affected by the
parameter variations, without making any reference to SSB. To investigate
this behavior, it is important to quantify the intensity of the spatial
SSB presented by the GS solitons. Here, we use the asymmetry ratio,
defined as 
\begin{equation}
\Theta_{1,2}=\frac{\int_{0}^{+\infty}|\phi_{1,2}|^{2}dx-\int_{-\infty}^{0}|\phi_{1,2}|^{2}dx}{P_{1,2}},\label{ASSY}
\end{equation}
where $\Theta$ has values from $-1$ to $1$. Asymmetric GS solitons
are characterized by $|\Theta|>0$, while symmetric profiles have
$|\Theta|=0$. In particular, values of $|\Theta|$ close to limit
$1$ indicate that the GS soliton is confined by a single potential
well (\ref{Pot}), as shown in Fig. \ref{F2}(c).

In Fig. \ref{F4}, we present the same analysis as in Fig. \ref{F3},
but considering the influence of the parameters on the (a)symmetry
region. We observe that the SSB is present for both components when
$-2.1<g_{1}=g_{2}<1.8$. Similarly, when $g_{2}$ is fixed, the system
exhibits SSB for $g_{1}<2.5$. With $g_{1}$ fixed, the SSB region
for both components expands slightly with variations in $g_{2}$.
In general, we observe that the decrease in nonlinearities favors
the emergence of SSB. These results are presented in Figs. \ref{F4}(a-c).
The depth of the double wells, $U_{0}$, also plays an important role
in the symmetry of the GS solitons. Fig. \ref{F4}(d) shows the behavior
of $|\Theta|$ versus $U_{0}$, where we observe that only asymmetric
profiles are found for $U_{0}<1.52$. Under the same conditions, Fig.
\ref{F4}(e) demonstrates that strong coupling also promotes SSB,
restricting symmetric profiles to weak couplings ($|g_{1,2}|<-2.1$).
Finally, Fig. \ref{F4}(f) illustrates the influence of total power
on the symmetry of the GS solitons, clearly showing that an increase
in $P$ leads to asymmetric profiles. Our findings indicate that modifying
the LI parameter exclusively is insufficient to affect the symmetry
of the coupled GS solitons. Despite this, fractional diffraction influence
the GS solitons in complex ways and, therefore, is not discussed in
this analysis. We detail these behaviors below.

We begin by presenting the influence of fractional diffraction on
the symmetric GS solitons. As shown in Fig. \ref{F1}(f), a reduction
in LI leads to GS solitons with higher amplitudes, as detailed in
Fig. \ref{F5}(a). In addition, the tails of the symmetric GS solitons
undergo significant changes, as a reduction in LI causes them to extend
over longer distances (see Fig. \ref{F5}(b)).

In Fig. \ref{F6}, we detail the effects of fractional diffraction
on the asymmetric GS solitons. In this configuration, the decrease
in LI also leads to an increase in the amplitude of the GS solitons.
Unlike the symmetric case, when SSB is present, the decrease in LI
introduces a critical value $\alpha_{c}$, for which no GS profile
is obtained with $\alpha<\alpha_{c}$. For instance, in the configuration
shown in Fig. \ref{F6}(a) with coupling $g_{12}=-3$, all profiles
obtained with $\alpha<0.92$ are unstable delta-type solutions with
either $P_{1}$ or $P_{2}=0$. Therefore, there are certain limiting
values for LI ($\alpha_{c}$), for which model (\ref{EQ1}) does not
support GS solitons. Increasing the coupling intensity in such systems
can lead to the emergence of asymmetric GS solitons localized near
the center of the double-well potential, as shown in Fig. \ref{F6}(b).
These central asymmetric profiles exhibit a relatively small asymmetry
ratio compared to other asymmetric profiles. For example, considering
$g_{12}=-4$, the standard asymmetric profiles for component 2 exhibit
$|\Theta_{2}|\approx1$, while the central asymmetric profiles obtained
with $\alpha=1.2$ and $1.25$ show $|\Theta_{2}|=0.13$ and $0.33$,
respectively. Numerical simulations demonstrate that increasing the
coupling intensity also raises the critical values $\alpha_{c}$.
This behavior is investigated in Fig. \ref{F6}(c), where the nonlinear
pattern of the critical LI, $\alpha_{c}(g_{12})$, is observed.

Considering the influence of coupling associated with other parameters
on the symmetry of the GS solitons, we extend the investigation by
means of existence diagrams, where the regions of existence of symmetric,
asymmetric, and central asymmetric GS solitons are determined. Furthermore,
we investigate the region where the formation of GS solitons in both
components does not occur. The existence diagrams of $g_{1}=g_{2}$,
$g_{1}$, $g_{2}$, $U_{0}$, $P$ and $\alpha$ versus $g_{12}$
are displayed in Fig. \ref{F7}. The regions of existence for the
different GS solitons are depicted in Fig. \ref{F7}(a), based on
the variations in nonlinearities and the coupling between the components.
We observe that the symmetric profiles are favored in strongly decocting
systems in a weak coupling regime. On the other hand, the SSB appears
in the strong coupling regime. In general, the optical system tends
not to support the GS solitons in the self-focusing regime with weak
coupling. By fixing $g_{1}$ or $g_{2}$, similar results are found
for the asymmetry regions. However, in this configuration, the symmetry
region is drastically reduced. Notably, when $g_{1}$ is fixed, weak
coupling prevents the formation of GS solitons, even under a strong
self-interaction regime in component 2. These findings are presented
in Figs. \ref{F7}(b-c).

The diagram in the plane $(g_{12},U_{0})$, shown in Fig. \ref{F7}(d),
demonstrates that SSB is favored by strong coupling in systems with
deep potentials. For $U_{0}<2.5$, symmetric profiles are also observed.
In this configuration, central asymmetric profiles are limited to
strongly coupled systems associated with a shallow external potential.
The absence of GS solitons is observed in regimes of strong coupling
associated with a deep potential. In Fig. \ref{F7}(e), we present
the results from the diagram ($g_{12}$, $P$). As observed previously,
increasing the coupling strength favors SSB. In this configuration,
GS solitons are not supported in systems with low total power in a
weakly coupled regime. The diagram corresponding to the variation
of the LI ($g_{12}$, $\alpha$) produces different results (see Fig.
\ref{F7}(f)). In this configuration, we observe that delta-type profiles
are favored in systems with small LI in a strong self-coupling regime.
Note that these states are not present in the previous diagrams, showing
their connection with the fractional diffraction and the coupling
intensity between the components. SSB is observed in systems with
large LI values in a strong coupling regime. In the weak coupling
regime, symmetric profiles are found. In particular, a small region
where SG profiles are not found is observed for weakly coupled systems
obtained with small LI.

In the previous analyses, we presented the regions of existence for
symmetric, asymmetric, central asymmetric, and delta-type profiles.
An important question to investigate is the dynamic stability of these
different profiles. For real-time propagation, we considered the coupled
GS solitons as input profiles, but randomly perturbed them by $3\%$
of their amplitudes. In these configurations, dynamic stability against
small perturbations is confirmed when both components propagate without
exhibiting abrupt changes in their shapes or amplitudes over a distance
of $z=10000$. As previously shown, the delta-type profiles are unstable
and exhibit only one component with non-zero individual power. Extensive
numerical simulations demonstrate the stability of all symmetric and
asymmetric (standard) profiles investigated. Despite the random noise
introduced into the amplitudes, both profiles evolve without significant
modifications to their shapes. In Fig. \ref{F8}(a-b), we present
examples of the stable evolution of GS solitons for component 2. In
both cases, component 1 shows similar results, which are not detailed
here.

In contrast, the central symmetric GS solitons present unstable evolution.
The dynamics of these profiles present coherent oscillations around
the minimum of one of the potential wells ($\pm x_{0}$). In the case
of the central asymmetric profiles that initially present $\Theta_{1,2}>0$,
the oscillatory dynamics occurs in the direction of $x>0$. Similarly,
for those that present $\Theta_{1,2}<0$, the oscillations occur in
the opposite direction, as shown in Fig. \ref{F8}(c). We also investigated
the oscillatory dynamics of the central symmetric profiles by considering
the period of the initial oscillations ($\tau$) with variations in
the parameters $g_{1}$, $g_{2}$, $U_{0}$, $g_{12}$, $P$ and $\alpha$.
Figs. \ref{F9}(a-b) show the effects of $g_{1}$ and $g_{2}$ on
the oscillation period of the GS profiles corresponding to component
2. Generally, an increase in nonlinearities leads to a reduction in
the oscillation period. When considering the depth of the double-well
potential, we observe that the oscillation period decreases as the
potential depth increases, as illustrated in Fig. \ref{F9}(c). When
the coupling is varied (see Fig. \ref{F9}(d)), the oscillation period
is seen to increase under strong coupling conditions. Fig. \ref{F9}(e)
demonstrates that the oscillations of the optical system are sensitive
to an increase in the total power. This behavior arises due to the
relationship between the total power and the system's energy. Finally,
in Fig. \ref{F9}(f), we investigated the influence of fractional
diffraction on the oscillation period, observing that an increase
in LI leads to shorter oscillation periods.

\section{Conclusion \label{Sec4}}

We introduce a one-dimensional model of coupled f-NLSE with a double-well
potential applied to only one of the components. In this configuration,
we investigate GS solitons, where SSB was observed in both the actuated
field and the partner component, highlighting asymmetry induction
via coupling. Numerical simulations reveal the existence of both symmetric
and asymmetric profiles, initiated from a slightly asymmetric initial
condition. The individual powers and the intensity of the SSB are
significantly influenced by the nonlinearities, potential depth, and
nonlinear coupling. Generally, asymmetry is favored in self-focusing
(or weakly self-defocusing) systems with strong coupling. Variations
in fractional diffraction affect the amplitude and tails of the coupled
symmetric GS solitons, while in asymmetric profiles, fractional diffraction
influences the localization position and dynamic stability. Additionally,
we identified critical values that define the minimum LI necessary
to produce coupled GS solitons. Diagrams were employed to illustrate
the regions where different types of GS solitons exist and where the
model fails to support localized solutions. The stability of the GS
solitons was assessed through direct simulations. For unstable cases,
characterized solely by centrally asymmetric GS solitons, the oscillatory
dynamics were analyzed, showing the impact of various parameters on
the oscillation period of these profiles. This work aims to contribute
new insights into SSB in fractional systems, particularly by elucidating
the sparsely explored topic of half-trapped fractional systems.

\section*{Acknowledgments}

The author acknowledges the financial support of the Brazilian agencies
CNPq (\#405638/2022-1, \#306105/2022-5 and Sisphoton Laboratory-MCTI
\#440225/2021-3) and CAPES. This work was also performed as part of
the Brazilian National Institute of Science and Technology (INCT)
for Quantum Information (\#465469/2014-0).

\bibliographystyle{apsrev4-2}
\bibliography{Refs}

\end{document}